\documentclass[letterpaper,12pt]{article}
\usepackage{tabularx} % extra features for tabular environment
\usepackage{amssymb}
\usepackage{amsmath}
\usepackage{amsmath,bm}
\usepackage{verbatim}
\usepackage{mathtools}
\usepackage{graphicx} % takes care of graphic including machinery
\usepackage[margin=1in,letterpaper]{geometry} % decreases margins
\usepackage{cite} % takes care of citations
\usepackage[final]{hyperref} % adds hyper links inside the generated pdf file
\usepackage{comment}
\usepackage{slashed}

\hypersetup{
	colorlinks=true,       % false: boxed links; true: colored links
	linkcolor=blue,        % color of internal links
	citecolor=blue,        % color of links to bibliography
	filecolor=magenta,     % color of file links
	urlcolor=blue         
}
\usepackage{bbold}

\begin{document}
	
	\newcommand{\N}{\mathcal{N}}
	\newcommand{\M}{\mathcal{M}}
	\newcommand{\A}{\mathcal{A}}
	\newcommand{\B}{\mathcal{B}}
	\newcommand{\C}{\mathcal{C}}
	\newcommand{\D}{\mathcal{D}}
	\newcommand{\E}{\mathcal{E}}
	\newcommand{\F}{\mathcal{F}}
	\newcommand{\G}{\mathcal{G}}
	\newcommand{\K}{\mathcal{K}}
%	$\M, \N, \A, \B, \C, \D,\E,\F,\G, \K $
%% template şuradan alındı
%%templates>more accademical journal templates>physics templates>Lab sample reports for...
%\title{Two pairs of 4-component spinors for SL(4,C) and SL(4,C)SL(4,C)*=SO(3,1) connection}

\section*{Electromagnetic stress-energy tensor is a direct consequence of 3- and 6-dimensional left-and right-helical representations of the Lorentz group} 
%%%%%%%%%%%%%% %%%%%%%%%%
\bigskip 

M. A. Kuntman
%\footnote{makuntman@gmail.com}

\bigskip
\bigskip

%\author{M. A. Kuntman}

%%%%%%%%%%%%%%%%%%%%%%%%%%%%%%%%

%\footnote{Independent researcher, Ordu, Turkey.}\;
% and E. Kuntman
%\footnote{Independent researcher, Izmir, Turkey (email: ahkuntman@gmail.com).}}               

%\date{May 24 2019}

%\maketitle

%%%%%%%%%%%%%%%%%%%%%%%%%%%%%%%%%%%%%%%%%%%%%%%%%%%%%%%%%%%%%%%%

\bigskip

%%%%%%%%%%%%%  %%%%%%%%%%%%  %%%%%%%%%%%%
%%%%%%%%%%  %%%%%%%%%%%%%%%%%%%%%% %%%%%%%%
%%%%%%%%%%%%%%%%  %%%%%%%%%%%%%%%%%%%%%

\subsubsection*{Abstract}
We review 3-d reducible representation of the Lorentz group and introduce a 6-d irreducible representation tailored for transforming 6-d electromagnetic vector, and we show that the mixture of the density matrices associated with the left- and right-helical states is equivalent to the electromagnetic stress-energy tensor. 

%%%%%%%%%%%%%%%%% %%%%%%%%%%%% %%%%%%%%%
%%%%%%%%%%% %%%%%%%%%%% %%%%%%%%%%%%
%%%%%%%%%%%%%% %%%%%%%%%%%% %%%%%%%%%%%%%

\subsection*{Introduction}

There are several ways of transforming the electric and magnetic fields: 

1. First transform the 4-vector $A^\mu$
\begin{equation}
A^\mu \rightarrow A'^\mu = \Lambda^\mu_{\:\:\nu}A^\nu
\end{equation}
and then find $\vec{E}$ and $\vec{B}$ from their definitions
%%%%%%%%%%%%%
 \begin{equation}
 	\vec{E}=-\nabla \phi -\frac{\partial \vec{A}}{\partial t}, \qquad \vec{B}=\nabla \times \vec{A}
 \end{equation}
%%%%%%%%%%%%%%%%%%

2. Transform the Faraday tensor
\begin{equation}
	F^{\mu \nu}\rightarrow F'^{\mu \nu}=\Lambda^\mu_{\:
\: \sigma}\Lambda^\nu_{\:
\: \rho}F^{\sigma \rho}
\end{equation}
where
%%%%%%%%%%%%%%%%%
\begin{equation}
	F^{\mu \nu}=\begin{pmatrix}
		0&-E_x&-E_y&-E_z\\E_x&0&-B_z&B_y\\E_y&B_z&0&-B_x\\E_z&-B_y&B_x&0
	\end{pmatrix}
\end{equation}
%%%%%%%%%%%%%%%%%%

3. Transform the Riemann-Silberstein vector 

\begin{equation}
	\xi_L\rightarrow \xi_L'=S_L\xi_L
\end{equation}
where $S_L$ is an element of the complexified SO(3) group that corresponds to the left-helicity representation and acts on the Riemann-Silberstein vector $\xi_L$ that represents left-helicity state:
\begin{equation}
	\xi_L=\vec{E}-i\vec{B}
\end{equation}
%%%%%%%%%%%%%
Similarly, we have right-helicity representation of the complexified SO(3) group $S_R$ that acts on the right-helicity state:
\begin{equation}
	\xi_R\rightarrow \xi_R'=S_R\xi_R
\end{equation}
%%%%%%%%%%%%%%
In a certain representation of basis generators left and right are related with each other by complex conjugation
\begin{equation}
	\xi_R=\xi_L^*, \qquad S_R=S_L^*
\end{equation}
%%%%%%%%%%%%%%%%%%%%%%
It is worth noting that this is a reducible representation that can be expressed as $(1, 0)\oplus (0, 1)$. Full transformation can be written as
\begin{equation}
	\begin{pmatrix}
		S_L&0\\0&S_R
	\end{pmatrix}\begin{pmatrix}
	\xi_L\\ \xi_R
	\end{pmatrix}
\end{equation}
where each block is a $3\times 3$ matrix.
%%%%%%%%%%%%%%%%%%%%

4. Recently we have introduced an irreducible 6-d vector representation of the Lorentz group $\Lambda_6$ that can be decomposed into left- and right-helical representations. $\Lambda_6$ is especially tailored for transformation of the 6-d real electromagnetic vector $\vec{F}$:
%%%%%%%%%%%%%%%%%%%%
\begin{equation}
	\vec{F}\rightarrow \vec{F}'=\Lambda_6 \vec{F}, \qquad \text{where}\qquad \vec{F}=\begin{pmatrix}
		\vec{E}\\\vec{B}
	\end{pmatrix}
\end{equation}
%%%%%%%%%%%%%%
It can be shown that $\Lambda_6$ can be written as a matrix product of 6-d left- and right-representations: $\Lambda_6=\mathcal{H}_L \mathcal{H}_R$.
%%%%%%%%%%%%%%%%%%%%%%%%%%%%%%%%%%%%%
In the chosen basis
$\mathcal{H}_R=\mathcal{H}_L^*$.

Left-helical representation $\mathcal{H}_L$ acts on the 6-component left-helical state $\mathcal{F}_L$:
%%%%%%%%%%%%%%%%%%%%
\begin{equation}
	\mathcal{F}_L\rightarrow \mathcal{F}_L'=\mathcal{H}_L \mathcal{F}_L \qquad \text{where} \qquad \mathcal{F}_L=\begin{pmatrix}
		\xi_L\\i\xi_L
	\end{pmatrix}
\end{equation}
%%%%%%%%%%%%%%%%%
Similarly, right-helical representation $\mathcal{H}_R$ acts on the 6-component right-helical state $\mathcal{F}_R$:
%%%%%%%%%%%%%%%%%%%%
\begin{equation}
	\mathcal{F}_R\rightarrow \mathcal{F}_R'=\mathcal{H}_R \mathcal{F}_R \qquad \text{where} \qquad \mathcal{F}_R=\begin{pmatrix}
		\xi_R\\-i\xi_R
	\end{pmatrix}
\end{equation}
%%%%%%%%%%%%%
%%%%%%%%%%%%%%%%%%%
%%%%%%%%%%%%% FOOTNOTE %%%%%%%%%%%%%%%%%%
\footnote{We have $\mathcal{F}_L^{\:\dagger}\mathcal{F}_R=0$, but $\xi_L^{\:\:\dagger}\xi_R\neq0$ }. 
%%%%%%%%%%%%%%%%%%%%%%%%%%%
%%%%%%%%%%%%%%%%%%%%%%%%%%%%

Lie algebras of the representations mentioned in cases 1-4 are all the same, namely the Lorentz algebra. In order to be self contained detailed information is given in the Appendix.
%%%%%%%%%%%%%%%%%%%%%%%%%

In this note we will show that the electromagnetic stress-energy tensor is a direct consequence of the 3- and 6-dimensional representations. 

%%%%%%%%%%%%%%%%%%%%%%%%%%%%
%%%%%%%%%%%%  %%%%%%%%%%%%%%%%%%%%%  %%%%%%%%%%
%%%%%%%%  %%%%%%%%%%%  %%%%%%%%%%%%%%%%%%%%
%%%%%%%%%%%%%  %%%%%%%%%%%%%%%%%%%%%%%%%%%%

\subsection*{Electromagnetic stress-energy tensor} 

Electromagnetic stress-energy tensor is a traceless symmetric tensor and it is given by
\begin{equation}
	T^{00}=\frac{1}{2}(E^2+B^2), \quad T^{0i}=(\vec{E}\otimes\vec{B})_i, \quad T^{ij}=-(E_iE_j+B_iB_j)+\frac{1}{2}(E^2+B^2)\delta^{ij}
\end{equation}
%%%%%%%%%%%%%%%%
It is straightforward to show that elements of this stress-energy tensor directly emerge from the tensor product of the helical states:  
%%%%%%%%%%%%%%
\begin{equation}
	\mathcal{T}=
	\mathcal{F}_L
	\mathcal{F}_L^{\dagger}+
	\mathcal{F}_R
	\mathcal{F}_R^{\dagger}
\end{equation}
%%%%%%%%%%%%%%%%%%%%
which transforms as
\begin{equation}
	\mathcal{T}\rightarrow \mathcal{T}'=\mathcal{H}_L(\mathcal{F}_L\mathcal{F}_L^{\dagger})\mathcal{H}^{\dagger}_L+\mathcal{H}_R(\mathcal{F}_R\mathcal{F}_R^{\dagger})\mathcal{H}^{\dagger}_R
\end{equation}
%%%%%%%%%%%%%%%%%%%%%%%%
%%%%%%%%%%%%%%%% FOOTNOTE %%%%%%%%%%%%%
\footnote{Since $
	\mathcal{F}_R=
	\mathcal{F}_L^*$, \quad
	%%%%%%%%%%%%%%%%%%%%%%%
	$
		\mathcal{T}=2\text{Re}(
		\mathcal{F}_L
		\mathcal{F}_L^{\dagger})$
	
}
%%%%%%%%%%%%%%%%%%%
Explicit form of $\mathcal{T}$ is given by
%%%%%%%%%%%%%%%%%%%%%%%%%%%%%  

\begin{equation}\footnotesize\label{T}
	\mathcal{T}=\begin{pmatrix}
	E_1^2+B_1^2&E_1E_2+B_1B_2&E_1E_3+B_1B_3&0&(\vec{E}\times \vec{B})_3&-(\vec{E}\times \vec{B})_2\\
	E_2E_1+B_2B_1&E_2^2+B_2^2&E_2E_3+B_2B_3&-(\vec{E}\times\vec{B})_3&0&(\vec{E}\times\vec{B})_1\\
	E_3E_1+B_3B_1&E_3E_2+B_3B_2&E_3^2+B_3^2&(\vec{E}\times\vec{B})_2&-(\vec{E}\times\vec{B})_1&0\\0&-(\vec{E}\times\vec{B})_3&(\vec{E}\times\vec{B})_2&E_1^2+B_1^2&E_1E_2+B_1B_2&E_1E_3+B_1B_3\\
	(\vec{E}\times\vec{B})_3&0&-(\vec{E}\times\vec{B})_1&E_2E_1+B_2B_1&E_2^2+B_2^2&E_2E_3+B_2B_3\\
	-(\vec{E}\times\vec{B})_2&(\vec{E}\times\vec{B})_1&0&E_3E_1+B_3B_1&E_3E_2+B_3B_2&E_3^2+B_3^2
	\end{pmatrix}
\end{equation}
%%%%%%%%%%%%%%%%%%%%%%%%%
$\mathcal{T}$ is neither symmetric nor traceless, but it is equivalent to $T^{\mu\nu}$. All elements of $T^{\mu\nu}$ can be directly read off from the elements of $\mathcal{T}$. For example, tr$\mathcal{T}$ gives $T^{00}$, etc. Here we write $T^{\mu\nu}$ in matrix form for reference:
%%%%%%%%%%%%%%%%%%%%%%%%%%
\begin{equation}\footnotesize
	T^{\mu\nu}=\begin{pmatrix}
		\frac{1}{2}(E^2+B^2)&(\vec{E}\times \vec{B})_1&(\vec{E}\times \vec{B})_2&(\vec{E}\times \vec{B})_3\\
		
		(\vec{E}\times \vec{B})_1&\frac{1}{2}(E^2+B^2)-(E_1^2+B_1^2)&-(E_1E_2+B_1B_2)&-(E_1E_3+B_1B_3)\\
		
		(\vec{E}\times\vec{B})_2&-(E_2E_1+B_2B_1)&\frac{1}{2}(E^2+B^2)-(E_2^2+B_2^2)&-(E_2E_3+B_2B_3)\\
		
		(\vec{E}\times\vec{B})_3&-(E_3E_1+B_3B_1)&-(E_3E_2+B_3B_2)&\frac{1}{2}(E^2+B^2)-(E_3^2+B_3^2)
	\end{pmatrix}
\end{equation}
%%%%%%%%%%%%%%%%%%%%%%%%%%%%%%%%%
%%%%%%%%%%%%%%%%%%%%%%%%%%%%%%%%%%%%%

%%%%%%%%%%%%%%%%%%%%%

%%%%%%%%%%%%%%%%%%%%%%%%%%%%%%
%%%%%%%%%%%%%%%%%%%%%%%%%%%%%%%%%%%%%
%%%%%%%%%%%%%%%%%% %%%%%%%%%%%%%%%%%%%%%
%%%%%%%%%%%%%%%%%%%%%%%%%%%%%%%%%%%%%%%%%%%

Equivalently, It is also possible to obtain the electromagnetic stress-energy tensor directly from the Riemann-Silbestein vectors: 
%%%%%%%%%%%%%%%%%
\begin{equation}\footnotesize\label{LL}
	\xi_L\xi_L^{\:\dagger}
	=\begin{pmatrix}
		E_1^2+B_1^2&E_1E_2+B_1B_2&E_1E_3+B_1B_3\\

		E_2E_1+B_2B_1&E_2^2+B_2^2&E_2E_3+B_2B_3\\
		
		E_3E_1+B_3B_1&E_3E_2+B_3B_2&E_3^2+B_3^2
	\end{pmatrix}+i\begin{pmatrix}
	0&(\vec{E}\times\vec{B})_3&-(\vec{E}\times\vec{B})_2\\

	-(\vec{E}\times\vec{B})_3&0&(\vec{E}\times\vec{B})_1\\
	
	(\vec{E}\times\vec{B})_2&-(\vec{E}\times\vec{B})_1&0
	\end{pmatrix}
\end{equation}
%%%%%%%%%%%%%%%%%%%%%%%%%%%
$\xi_L\xi_L^{\:\dagger}$ represents a pure state. Imaginary part gives the Poynting vector. All elements of the stress-energy tensor $T^{\mu\nu}$ can be read from $\xi_L\xi_L^{\:\dagger}$. Similarly we write the density matrix corresponding to the pure state $\xi_R\xi_R^{\:\dagger}$: 
%%%%%%%%%%%%%%%%%%%%%%%%%%%
\begin{equation}\footnotesize
	\xi_R\xi_R^{\:\dagger}
	=\begin{pmatrix}
		E_1^2+B_1^2&E_1E_2+B_1B_2&E_1E_3+B_1B_3\\

		E_2E_1+B_2B_1&E_2^2+B_2^2&E_2E_3+B_2B_3\\
		
		E_3E_1+B_3B_1&E_3E_2+B_3B_2&E_3^2+B_3^2
	\end{pmatrix}-i\begin{pmatrix}
		0&(\vec{E}\times\vec{B})_3&-(\vec{E}\times\vec{B})_2\\

		-(\vec{E}\times\vec{B})_3&0&(\vec{E}\times\vec{B})_1\\
		
		(\vec{E}\times\vec{B})_2&-(\vec{E}\times\vec{B})_1&0
	\end{pmatrix}
\end{equation}
%%%%%%%%%%%%%%%%%%%%%
 Now write the mixture of states 
 $\xi_L\xi_L^{\:\dagger}$ and $\xi_R\xi_R^{\:\dagger}$:
%%%%%%%%%%%%%%%%%%%%%%
\begin{equation}
	\xi_L\xi_L^{\:\dagger}+\xi_R\xi_R^{\:\dagger}
	=2\begin{pmatrix}
		E_1^2+B_1^2&E_1E_2+B_1B_2&E_1E_3+B_1B_3\\		
		E_2E_1+B_2B_1&E_2^2+B_2^2&E_2E_3+B_2B_3\\				E_3E_1+B_3B_1&E_3E_2+B_3B_2&E_3^2+B_3^2
	\end{pmatrix}
 \end{equation}
%%%%%%%%%%%%%%%%%%
%%%%%%%%%%%%%%%%%%%%
We get a density matrix of a mixture with no Poynting vector.

%%%%%%%%%%%%%%%%%%%%%%%%%%%%%%%%%%%%%%%%%%%%
%%%%%%%%%%%%%%%%%  %%%%%%%%%%%%%%%%%%%%%%%%%
%%%%%%%%%%% %%%%%%%%%%%%%%%%%% %%%%%%%%%%%%%%
%%%%%%%%%%%%%%%%%%%%%%%%%%%%%%%%%%%%%%%%%%%%%
\subsection*{Appendix}
%%%%%%%%%%%%%%%%%%
%%%%%%%%%%%%%%%%%
%%%%%%%%%%%%%%%%%%%%%%
%%%%%%%%%%%%%%%%%%%%%%

%%%%%%%%%%%%%%%%%%%%%%%%%%

\subsubsection*{1. (1, 0)$\oplus$ (0, 1) reducible representation of the Lorentz group}

Generators of the complexified SO(3) group:
%%%%%%%%%%%
\begin{equation}\footnotesize
	 J_1=\begin{pmatrix}
		0&0&0\\0&0&-1\\0&1&0
	\end{pmatrix} , \quad J_2= \begin{pmatrix}
		0&0&1\\0&0&0\\-1&0&0
	\end{pmatrix} ,\quad J_3= \begin{pmatrix}
		0&-1&0\\1&0&0\\0&0&0
	\end{pmatrix}  
\end{equation}
%%%%%%%%%%%%%%%%%%%%%%%
\begin{equation}\footnotesize
	K_1=\begin{pmatrix}
		0&0&0\\0&0&-i\\0&i&0
	\end{pmatrix} , \quad K_2= \begin{pmatrix}
		0&0&i\\0&0&0\\-i&0&0
	\end{pmatrix} ,\quad K_3= \begin{pmatrix}
		0&-i&0\\i&0&0\\0&0&0
	\end{pmatrix}  
\end{equation}
%%%%%%%%%%%%%%%%%%%%%%%
We have the following commutation relations:
\begin{equation}
	[J_i, J_j]=\epsilon_{ijk}J_k, \qquad [K_i, K_j]=-\epsilon_{ijk} J_k, \qquad [J_i, K_j]=\epsilon_{ijk}K_k	
\end{equation}
%%%%%%%%%%%%%%%%%%%%%%%%%%%%%%%%
This is the algebra of the Lorentz group.
%%%%%%%%%%%%%%%%%%%%%%%%%%%%
In terms of these generators we can write $S_L$ and $S_R$ as follows:
%%%%%%%%%%%%%%%%%%%%%%%%
\begin{equation}
	S_L=e^{(\theta_i+i\phi_i) J_i}, \qquad S_R=e^{(\theta_i-i\phi_i) J_i} 
\end{equation}
%%%%%%%%%%%%%%%%%%%%%%
where $\theta_i$ and $\phi_i$ are the rotation and boost parameters respectively.
%%%%%%%%%%%%%%%%%%%%%%
%%%%%%%%%%%%%%%%%%%%%%%
 
It is worth noting that this is a reducible representation that can be expressed as $(1, 0)\oplus (0, 1)$. Full transformation can be written as
\begin{equation}
	\begin{pmatrix}
		S_L&0\\0&S_R
	\end{pmatrix}\begin{pmatrix}
		\xi_L\\ \xi_R
	\end{pmatrix}
\end{equation}
each block is a $3\times 3$ matrix.
%%%%%%%%%%%%%%%%%%%%

%%%%%%%%%%%%%%%%%
%%%%%%%%%%%%%%
%%%%%%%%%%%%%%%%%%%%%%%%%

%%%%%%%%%%%%%%%%%%%%%%%%%%%

%%%%%%%%%%%%%%%%%%%%%%%%%%%%
%%%%%%%%%%%%%%%%%%%%%%%%%%%%

%%%%%%%%%%%%%%%%%
%%%%%%%%%%%%%%%  %%%%%%%%%%%%%%%%%%%%%%%%
%%%%%%%%%%%%%%   %%%%%%%%%%%%%%%%%%%%%%%

%%%%%%%%%%%%%%%%%%%%%%%%%%%%%%%%%%%%%%%%%%%%%%%%%%%%%%%%%%%%%%%%%%%%%%%%%%%%%%%%%%%%%%%%%%%%%%%%%%%%%%%%%%%%%%%%%%%%

%%%%%%%%%%%%%%%% %%%%%%%%%%%%%% %%%%%%%%
%%%%%%%%%%% %%%%%%%%%%%%  %%%%%%%%%%%

%%%%%%%%%%%%%%%%%%%%%%%%%%%%%%%%%%%%%%%
%%%%%%%%%%%%%%%%%%%%%%%%%%%%%%%%%%

%%%%%%%%%%%%%%%%%%%%%%%%%%%%%%%%%
%%%%%%%%%%%%%%%%%%  %%%%%%%%%%%%%%%%%%%%%%%%
%%%%%%%%%%  %%%%%%%%%%%%%%%%%  %%%%%%%%%%%%
\subsubsection*{2. Six-dimensional irreducible representations of the Lorentz group}
\subsubsection*{2.1. Extension and complexification of the group SO(3)}

Let $\vec{F}$ be the 6-component electromagnetic vector:
%%%%%%%%%%%%%%
\begin{equation}
	\vec{F}=\begin{pmatrix}\vec{E}\\\vec{B}	\end{pmatrix}	
\end{equation}
%%%%%%%%%%%%%%%%%%%
And let $\Lambda_6$ be a $6\times6$ Lorentz transformation matrix acting on $\vec{F}$ that can be written in terms of parameters $\theta_i$ and $\phi_i$   and  rotation and boost generators  $J_i$ and $K_i$:
\begin{equation}
	\Lambda_6=e^{(\theta_i J_i+ \phi_i K_i)}
\end{equation}
%%%%%%%%%%%%
where
%%%%%%%%%%%%%%%%%%%%%
\begin{equation}
	J_1=e\otimes S_1, \qquad J_2=e\otimes S_2, \qquad J_3=e\otimes S_3
\end{equation}
\begin{equation}
	K_1=\epsilon\otimes S_1, \qquad K_2=\epsilon\otimes S_2, \qquad K_3=\epsilon\otimes S_3
\end{equation}
%%%%%%%%%%%%%%%%%%%%%%%%%%%%
$S_1, S_2, S_3$ are the generators of the SO(3) group and $e$ and $\epsilon$ are the extension units 
with the properties, $e^2=\mathbb{1}_{2\times 2}$, $\epsilon^2=-\mathbb{1}_{2\times 2}$:
\begin{equation}\footnotesize \label{S}
	e=\begin{pmatrix}
	1&0\\0&1
	\end{pmatrix}, \quad \epsilon=\begin{pmatrix}
	0&1\\-1&0
	\end{pmatrix},\quad S_1=\begin{pmatrix}
	0&0&0\\0&0&-1\\0&1&0
	\end{pmatrix} , \quad S_2= \begin{pmatrix}
	0&0&1\\0&0&0\\-1&0&0
	\end{pmatrix} ,\quad S_3= \begin{pmatrix}
	0&-1&0\\1&0&0\\0&0&0
	\end{pmatrix}  
\end{equation}
%%%%%%%%%%%%%
%%%%%%%%%%%%%%%%%%%%%%%
In this irreducible representation $\epsilon$ plays the role of the imaginary unit and serves for the complexification of the algebra.
%%%%%%%%%%%%%%%%%%%%%%%%%%%%%%%
%%%%%%%%%%%%%%%%%%%%%%%%%

%%%%%%%%%%%%%%%%%%%%%%%%%%%

We have the following commutation relations:
\begin{equation}
	[J_i, J_j]=\epsilon_{ijk}J_k, \qquad [K_i, K_j]=-\epsilon_{ijk} J_k, \qquad [J_i, K_j]=\epsilon_{ijk}K_k	
\end{equation}
%%%%%%%%%%%%%%%%%%%%%%%%%%%%%%%%
This is the algebra of SO(1,3).
%%%%%%%%%%%%%%%%%%%%%%%%%%%%
%%%%%%%%%%%%%%%%%%%%%%%%%%%%

Under the rotations electric and magnetic field vectors rotate independently. Boosts are more interesting. Explicit form of a boost in an arbitrary direction and its action on $\vec{F}$ reads
%%%%%%%%%%%%%%%%%%%%%%
\begin{equation}
	\begin{pmatrix}
		E'_x\\E'_y\\E'_z\\B'_x\\B'_y\\B'_z
	\end{pmatrix}=\begin{pmatrix}
	\gamma+\frac{\beta_x^2(1-\gamma)}{\beta^2}&\frac{\beta_x\beta_y(1-\gamma)}{\beta^2}&\frac{\beta_x\beta_z(1-\gamma)}{\beta^2}&0&-\beta_z\gamma&\beta_y\gamma\\
	
	\frac{\beta_y\beta_x(1-\gamma)}{\beta^2}&\gamma+\frac{\beta_y^2(1-\gamma)}{\beta^2}&\frac{\beta_y\beta_z(1-\gamma)}{\beta^2}&\beta_z\gamma&0&-\beta_x\gamma\\
	
	\frac{\beta_z\beta_x(1-\gamma)}{\beta^2}&\frac{\beta_z\beta_y(1-\gamma)}{\beta^2}&\gamma+\frac{\beta_z^2(1-\gamma)}{\beta^2}&-\beta_y\gamma&\beta_x\gamma&0\\
	
	0&\beta_z\gamma&-\beta_y\gamma&\gamma+\frac{\beta_x^2(1-\gamma)}{\beta^2}&\frac{\beta_x\beta_y(1-\gamma)}{\beta^2}&\frac{\beta_x\beta_z(1-\gamma)}{\beta^2}\\
	
	-\beta_z\gamma&0&\beta_x\gamma&\frac{\beta_y\beta_x(1-\gamma)}{\beta^2}&\gamma+\frac{\beta_y^2(1-\gamma)}{\beta^2}&\frac{\beta_y\beta_z(1-\gamma)}{\beta^2}\\
	
	\beta_y\gamma&-\beta_x\gamma&0&\frac{\beta_z\beta_x(1-\gamma)}{\beta^2}&\frac{\beta_z\beta_y(1-\gamma)}{\beta^2}&\gamma+\frac{\beta_z^2(1-\gamma)}{\beta^2}
	\end{pmatrix} \begin{pmatrix}
	E_x\\E_y\\E_z\\B_x\\B_y\\B_z
	\end{pmatrix}
\end{equation}
%%%%%%%%%%%%%%%%%
%%%%%%%%%%%%%%%%%%%%%%%

The irreducible representation $\Lambda_6$ preserves a metric $g$ which plays the role of the Minkowski metric:
\begin{equation}\label{Met}
   \Lambda_6^Tg\Lambda_6=g , \quad \text{where},\quad g=\begin{pmatrix}
   	1&0&0&0&0&0\\0&1&0&0&0&0\\0&0&1&0&0&0\\0&0&0&-1&0&0\\0&0&0&0&-1&0\\0&0&0&0&0&-1
   \end{pmatrix}
\end{equation}
%%%%%%%%%%%%%%%%%%%%%%%
%%%%%%%%%%%%%%%%%%
With this metric we can define the scalar product and obtain one of the Lorentz invariants of the electromagnetic field:   
%%%%%%%%%%%%%%%%%%%%%%%%%%
\begin{equation}
	\vec{F}\cdot\vec{F}= \vec{F}^Tg\vec{F}=|\vec{E}|^2-|\vec{B}|^2
\end{equation}
%%%%%%%%%%%%%%%%%%%%%%%%%%%%%%%%%%%
%%%%%%%%%%%%%%%
%%%%%%%%%%%%%%%%%%%%%%%%%%otimes 
%%%%%%%%%%%%%%%  %%%%%%%%%%%%%%%%%%%%%

%%%%%%%%%%  %%%%%%%%%%%%%%%%%%%%%%%
%%%%%%%%%%%%%%%%%%%  %%%%%%%%%%%%%%%%

\subsubsection*{3.2. Left- and right-helical representations}

Let us define new generators:
\begin{equation}\footnotesize    M_i=\frac{1}{2}(J_i-iK_i)=\frac{1}{2}(e\otimes S_i-i\epsilon\otimes S_i)=h\otimes S_i, \qquad N_i=\frac{1}{2}(J_i+iK_i)=\frac{1}{2}(e\otimes S_i+i\epsilon\otimes S_i)=h^*\otimes S_i 
\end{equation}
%%%%%%%%%%
where
\begin{equation}
	h=\frac{1}{2}\begin{pmatrix}
		1&-i\\i&1
	\end{pmatrix}, \qquad h^n=h,\qquad hh^*=0	
\end{equation}

$M_i$ and $N_i$ obey the SU(2) algebra and commute with each other. We rewrite $\Lambda_6$ in terms of $M_i$ and $N_i$ or equivalently in terms of $h\otimes S_i$ and $h^*\otimes S_i$:
%%%%%%%%%%%%%%%%%%
 \begin{equation}\label{L6}
	\Lambda_6=e^{(\theta_i+i\phi_i)h\otimes S_i+(\theta_i-i\phi_i)h^*\otimes S_i}
\end{equation}
%%%%%%%%%%%%%%%%%%%%%%%
%%%%%%%%%%%%%%%%%%%%
Let us denote $\mathcal{H}_L=e^{(\theta_i+i\phi_i)h\otimes S_i}$ and $\mathcal{H}_R=e^{(\theta_i-i\phi_i)h^*\otimes S_i}$. Since $[M_i, N_j]=0$, $\Lambda_6$ can be written as matrix product, $\Lambda_6=\mathcal{H}_L\mathcal{H}_R$, where $\mathcal{H}_L$ and $\mathcal{H}_R$ can be regarded as 6-dimensional left- and right-helical irreducible representations. $\mathcal{H}_L$ acts on the left-helical state $\mathcal{F}_L$
%%%%%%%%%%%%%%%%%
%%%%%%%%%%% FOOTNOTE %%%%%%%%%%%%%%%%%%%
\footnote{In this representation $\mathcal{H}_R=\mathcal{H}_L^*$, but this is not true in general.}:
%%%%%%%%%%%%%%%%%%%%%%%%%%
%%%%%%%%%%%%%%%%%%%%%%%%

%%%%%%%%%%%%%%%%}
\begin{equation}
\mathcal{F}_L = \begin{pmatrix}
	\mathcal{E}\\i\mathcal{E}
\end{pmatrix},\qquad \text{where}\qquad \mathcal{E}=\vec{E}-i\vec{B}
\end{equation}
%%%%%%%%%%%%%%%%%%%%%%%%0&
%%%%%%%%%%%%%%%%%%%%%%%%%%%%

%%%%%%%%%%%%%%%%%%%%%%%%%%%%%
%%%%%%%%%%%%%%%%%%%%%%%%%%%%%%%%

As an example let us write the boost matrix in the $x$-direction for $ \mathcal{H}_L$ 
%%M%%%%%%%% LEFT BOOST MATRİX %%%%%%%%%%%%%%%%%%%%
\begin{equation}
\mathcal{H}_L=
\begin{pmatrix}
	1&0&0&0&0&0\\
	0&\frac{1}{2}(1+\cosh\phi)&-\frac{i}{2}\sinh\phi&0&\frac{i}{2}(1-\cosh\phi)&-\frac{1}{2}\sinh\phi
	\\0&\frac{i}{2}\sinh\phi&\frac{1}{2}(1+\cosh\phi)&0&\frac{1}{2}\sinh\phi&\frac{i}{2}(1-\cosh\phi)\\
	0&0&0&1&0&0\\
	0&-\frac{i}{2}(1-\cosh\phi)&\frac{1}{2}\sinh\phi&0&\frac{1}{2}(1+\cosh\phi)&-\frac{i}{2}\sinh\phi\\
	0&-\frac{1}{2}\sinh\phi&-\frac{i}{2}(1-\cosh\phi)&0&\frac{i}{2}\sinh\phi&\frac{1}{2}(1+\cosh\phi)
	
\end{pmatrix} 
\end{equation}
%%%%%%%%%%%%%%%%%%%%%%%%%%%%%%%%%%%%%%%%
%%%%%%
Similar expressions can be written for the right-helical representation.
%%%%%%%%%%%%%%%%%%%%%%%%%%%%%%%
%%%%%%%%%%%%%%%%%%%%%%%%%%%%%%%%

The  metric $g$ that given in Eq.\eqref{Met} is also preserved by 
$\mathcal{H}_L$ and $\mathcal{H}_R$:
%%%%%%%%%%%%%
\begin{equation}
\mathcal{H}_L^Tg\mathcal{H}_L=g, \qquad \mathcal{H}_R^Tg\mathcal{H}_R=g 
\end{equation}
%%%%%%%%%%%%%%%%%%
Hence we can define the scalar product of helical states and find the associated Lorentz invariants:
%%%%%%%%%%%%%%5
\begin{equation}
\mathcal{F}_L\cdot \mathcal{F}_L=\mathcal{F}^T_L g \mathcal{F}_L=|\vec{E}|^2-|\vec{B}|^2-2i\vec{E}\cdot\vec{B}
\end{equation}
%%%%%%%%%%%%%%%%%%%%%
%%%%%%%%%%%%%%%%  %%%%%%%%%%%%%%%%%%%
%%%%%%%%%%  %%%%%%%%%%%%%%%%%  %%%%%%%%%%%%%
%%%%%%%%%%%%%%%%%%%%%%%%%%%%%%%%%%%%%%%%%

\subsubsection*{3.2 Weyl-like equations for left- and right-helical states}

Let $\mathbf{\Sigma}=i(S_x, S_y, S_z)$, where $S_i$ is given in Eq.\eqref{S}, and let us define the operator $\mathbf{\Sigma}\cdot\mathbf{n}$:
%%%%%%%%%%%%%%%%%%%%%%%%%%%
\begin{equation}
	\mathbf{\Sigma}\cdot\mathbf{n} =\begin{pmatrix}
		0&-ic&is\slashed{s}\\ic&0&-is\slashed{c}\\-is\slashed{s}&is\slashed{c}&0\end{pmatrix}	
	\end{equation}
%%%%%%%%%%%%%%
where $\mathbf{n}=(\sin\theta\cos\phi, \sin\theta\sin\phi, \cos\theta)$ and $c=\cos\theta,\: s=\sin\theta, \: \slashed{c}=\cos\phi, \: \slashed{s}=\sin\phi$. $\mathbf{\Sigma}\cdot\mathbf{n}$ is a traceless Hermitian matrix and it has 3 orthogonal eigenvectors corresponding to the eigenvalues $\pm 1$ and $0$.

\begin{equation}\label{Weyl}
	\varphi_+=\frac{1}{\sqrt{2}}\begin{pmatrix}-c\slashed{c}+i\slashed{s}\\-c\slashed{s}-i\slashed{c}\\s
	\end{pmatrix}, \qquad \varphi_-=\frac{1}{\sqrt{2}}\begin{pmatrix}-c\slashed{c}-i\slashed{s}\\-c\slashed{s}+i\slashed{c}\\s
	\end{pmatrix}, \qquad \varphi_0=\begin{pmatrix}s\slashed{c}\\s\slashed{s}\\c
	\end{pmatrix}
\end{equation} 
%%%%%%%%%%%%%%%%%%
%%%%%%%%%%%%%%%%%%%%%%%%%%%%%%%
%%%%%%%%%%%%%%%%%%%%%%%%%%%%%%%%%
Actually, these are 3-component objects similar to the Weyl spinors, and they are the solutions to the following version of the Weyl equations:  
%%%%%%%%%%
\begin{equation}\label{ww2}
	(i\frac{\partial }{\partial t}-\mathbf{\Sigma}\cdot\mathbf{p})\psi=0 
\end{equation}
\begin{equation}\label{ww3}
	(i\frac{\partial }{\partial t}+\mathbf{\Sigma}\cdot\mathbf{p})\chi=0
\end{equation}
%%%%%%%%%%%%
where $\mathbf{p}=|\mathbf{p}|\mathbf{n}$. Plane wave ansatz suggest that $\varphi_+$ and $\varphi_-$ are the solutions that can be associated with right- and left-helicity states.
%%%%%%%%%%%%%%%%%%%%%
%%%%%%%%%%%%%%%%%%%%%%

We can also make the connection between $\varphi_{\pm}$ and $\xi_{R,L}$. Let us start with Eq.\eqref{ww2} and suppose that E=$|\mathbf{p}|>0$ and assume that $\xi_L$ satisfies the equation:
%%%%%%%%%%%%%%%%%%%%%%%%
\begin{equation}
\begin{pmatrix}
	-1&-ic&is\slashed{s}\\ic&-1&-is\slashed{c}\\-is\slashed{s}&is\slashed{c}&-1\end{pmatrix}
	\begin{pmatrix}
		E_x-iB_x\\E_y-iB_y\\E_z-iB_z
	\end{pmatrix}	=0
\end{equation}
%%%%%%%%%%
This matrix equation gives 4 linearly independent equations to be solved for $E_i$ and $B_i$. Taking $E_3$ and $B_3$ as free variables we get the following expressions:
\begin{equation}\footnotesize
	E_1=(-c\slashed{c}E_3-\slashed{s}B_3)/s, \quad
E_2=(-c\slashed{s}E_3+\slashed{c}B_3)/s, \quad B_1=(-c\slashed{c}B_3+\slashed{s}E_3)/s, \quad B_2=(-c\slashed{s}B_3-\slashed{c}E_3)/s
\end{equation}
%%%%%%%%%%%%%%%%%%%%%%%
Hence, $\xi_L$ can be written as
\begin{equation}\xi_L=\begin{pmatrix}
		E_x-iB_x\\E_y-iB_y\\E_z-iB_z
	\end{pmatrix}	=\frac{E_3-iB_3}{s}\begin{pmatrix}
	-c\slashed{c}-i\slashed{s}\\-c\slashed{s}+i\slashed{c}\\s
	\end{pmatrix}=\frac{E_3-iB_3}{s}\varphi_L
\end{equation}
%%%%%%%%%%%%%%%%%%%%
Similar expression can be written for $\xi_R$ by simply taking the complex conjugate of $\xi_L$.
%%%%%%%%%%%%%%%%%%%%%%%%%%%%%%%%%%%
\begin{equation}\xi_R=\begin{pmatrix}
		E_x+iB_x\\E_y+iB_y\\E_z+iB_z
	\end{pmatrix}	=\frac{E_3+iB_3}{s}\begin{pmatrix}
		-c\slashed{c}+i\slashed{s}\\-c\slashed{s}-i\slashed{c}\\s
	\end{pmatrix}=\frac{E_3+iB_3}{s}\varphi_R
\end{equation}

%%%%%%%%%%%%%%  %%%%%%%%%%%%%%%%%%%%%%%%%%%%
\subsubsection*{3.3 Duals and their transformations}

Let $\mathcal{\widetilde{F}}_L$ and $\mathcal{\widetilde{F}}_R$ be the duals of $\mathcal{F}_L$ and $\mathcal{F}_R$:
\begin{equation}
	\mathcal{\widetilde{F}}_L=g \mathcal{F}_L=\begin{pmatrix}
		\mathcal{E}_L\\-i\mathcal{E}_L
	\end{pmatrix}, \qquad  \mathcal{\widetilde{F}}_R=g \mathcal{F}_R=\begin{pmatrix}
	\mathcal{E}_R\\i\mathcal{E}_R
	\end{pmatrix},
	 \quad \text{where}\quad \mathcal{\widetilde{F}}_R=\mathcal{\widetilde{F}}^*_L
\end{equation}
%%%%%%%%%%%%%%%%
It can be shown that $\mathcal{\widetilde{F}}_R$ transforms as follows:
\begin{equation}
	\mathcal{\widetilde{F}}_R\rightarrow \mathcal{\widetilde{F}}'_R=(\mathcal{H}^{-1}_L)^{\dagger}\mathcal{\widetilde{F}}_R
\end{equation}
Here we have the property:
\begin{equation}
	(\mathcal{H}^{-1}_L)^{\dagger}=g\mathcal{H}^*_L g^{-1}
\end{equation}
%%%%%%%%%%%%%%%%%%%
Similarity to the Weyl spinors and their transformation transformation properties can be made more clear by using the Infeld-Van der Wearden notation:
%%%%%%%%%%%%%%
\begin{equation}
	\mathcal{F}_L=\mathcal{F}^*_R=\mathcal{F}_a, \qquad \mathcal{F}^*_L=\mathcal{F}_R=\mathcal{F}_{\dot{a}}, \qquad \widetilde{\mathcal{F}}_L=\widetilde{\mathcal{F}}_R^*=\mathcal{F}^a, \qquad \widetilde{\mathcal{F}}^*_L=\widetilde{\mathcal{F}}_R=\mathcal{F}^{\dot{a}}    
\end{equation}
%%%%%%%%%%%%%%%%%%%%%%%%%%%%%%%%%%%%%%%%%%%
%%%%%%%%%%%%%%%%%%%%%%%%%%%%%%%%%%%%%%%%%%%%
%%%%%%%%%%%%%%%% %%%%%%%%%%%%%%%%%%%%%%%%%%%%

\subsubsection*{3.4 Parity transformation}

Under the parity $\vec{E}$ transforms as a vector and $\vec{B}$ transforms as an axial vector:
\begin{equation}
\vec{E}\xrightarrow{P} -\vec{E}, \qquad \vec{B}\xrightarrow{P}\vec{B}
\end{equation}
%%%%%%%%%%%%%%%%%%
Accordingly, under the parity inversion the Riemann-Silberstein vector transforms as
\begin{equation}
	\xi_L=\vec{E}-i\vec{B}\xrightarrow{P} -\vec{E}-i\vec{B}=-\xi^*_L=-\xi_R, \qquad
\end{equation}
\begin{equation}
	\xi_R=\vec{E}+i\vec{B}\xrightarrow{P} -\vec{E}+i\vec{B}=-\xi^*_R=-\xi_L, \qquad
\end{equation}
%%%%%%%%%%%%%%%%%%%%%
On the other hand, under the parity inversion left- and right-helical states behave in a different way:
\begin{equation}
	\mathcal{F}_L=\begin{pmatrix}
		\xi_L\\i\xi_L
	\end{pmatrix}\xrightarrow{P}\begin{pmatrix}
	-\xi^*_L\\-i\xi^*_L
	\end{pmatrix}=-\widetilde{\mathcal{F}}_R, \quad \text{similarly}  \quad 	\mathcal{F}_R\xrightarrow{P}-\widetilde{\mathcal{F}}_L	
\end{equation} 
%%%%%%%%%%%%%%%%%%%%
\begin{equation}
	\widetilde{\mathcal{F}}_L=\begin{pmatrix}
		\xi_L\\-i\xi_L
	\end{pmatrix}\xrightarrow{P}\begin{pmatrix}
		-\xi^*_L\\i\xi^*_L
	\end{pmatrix}=-\mathcal{F}_R \quad \text{similarly} \quad \widetilde{\mathcal{F}}_R\xrightarrow{P}-\mathcal{F}_L
\end{equation} 
%%%%%%%%%%%%%%%%%
It may be easier to see these transformation  properties of the helical states in the Infeld-Van der Wearden notation
\begin{equation}
	\mathcal{F}_a\xrightarrow{P}-\mathcal{F}^{\dot{a}}, \quad \mathcal{F}_{\dot{a}}\xrightarrow{P}-\mathcal{F}^{a} ,\quad
	\mathcal{F}^{a}\xrightarrow{P}-\mathcal{F}_{\dot{a}}, \quad \mathcal{F}^{\dot{a}}\xrightarrow{P}-\mathcal{F}_{a} 	
\end{equation} 
%%%%%%%%%%%%%%%%%%%%
Minus signs in these relations have no significance. We can get rid of them by simply redefining $g$ as $-g$.
%%%%%%%%%%%%%%%%%%%%%%%%%%%%%%%%%%%%%%%%%%%%%%
%%%%%%%%%%%%%%%  %%%%%%%%%%%%%%  %%%%%%%%%%%%%%%%%%%%%%%%%%%%%% g$ as $-g$. %%%%%%%%%%%%%%%%%%%%%%%  %%%%%%%%%%%%%%%%%%%%%%%%%%%% 

%%%%%%%%%%%%%%%%%%%%%%%%%%%%%%
%%%%%%%%%%%%%%%%  %%%%%%%%%%%%%%%%%%%%%%%%
%%%%%%%%%%% %%%%%%%%%%%%%%%%% %%%%%%%%%%%%%%
%%%%%%%%%%%%%%%%%%%  %%%%%%%%%%%%%%%%%%%

\end{document}